\documentclass[superscriptaddress,prb,preprint,amsmath,amssymb]{revtex4}

\usepackage{graphicx,natbib}

\usepackage{calc,graphicx,float,amsmath}

\usepackage[paper=letterpaper,
            includefoot,
            marginparsep=.05in,       
            margin=0.75in,               
            includemp]{geometry} 
            
\def\sex{ {s^{\textnormal{ex}}} }
\def\stwo{ {s^{(2)}} }
\def\stwoeff{ {s^{(2)}_{\textnormal{eff}}} }

\def\taueff{ {\tau_{\textnormal{eff}}} }
\def\geff{ g_{\textnormal{eff}} }
\usepackage{varioref}
\begin{document}

\title{Implications of the effective one-component analysis of pair correlations in colloidal fluids with polydispersity}

\author{Mark J. Pond} 
\affiliation{Department of Chemical Engineering,
  The University of Texas at Austin, Austin, TX 78712.}

\author{Jeffrey R. Errington} 
\affiliation{Department of Chemical and Biological Engineering,
  University at Buffalo, The State University of New York, Buffalo,
  New York 14260-4200, USA}

\author{Thomas M. Truskett} \email{truskett@che.utexas.edu}
\thanks{Corresponding Author} {}
\affiliation{Department of Chemical Engineering, The University of Texas
  at Austin, Austin, TX 78712.}

\begin{abstract}
Partial pair-correlation functions of colloidal suspensions with continuous polydispersity can be challenging to characterize from optical microscopy or computer simulation data due to inadequate sampling. As a result,  it is common to adopt an effective one-component description of the structure that ignores the differences between particle types. Unfortunately, whether this kind of simplified description preserves or averages out information important for understanding the behavior of the fluid depends on the degree of polydispersity and can be difficult to assess, especially when the corresponding multicomponent description of the pair correlations is unavailable for comparison. Here, we present a computer simulation study that examines the implications of adopting  an effective one-component structural description   of a polydisperse fluid. The square-well model that we investigate mimics
key aspects of the experimental behavior of suspended colloids with short-range, polymer-mediated
attractions. To characterize the partial pair-correlation functions and thermodynamic excess entropy of this system, we introduce a Monte Carlo sampling strategy appropriate for  fluids with a large number of pseudo-components. The data from our simulations at high particle concentrations, as well as exact theoretical results for dilute systems, show how qualitatively different trends between structural order and particle attractions  emerge from the multicomponent and effective one-component treatments, even with systems characterized by moderate polydispersity. We examine  consequences of these differences for excess-entropy based scalings of shear viscosity, and we discuss how use of the multicomponent treatment reveals similarities between the corresponding dynamic scaling behaviors of attractive colloids and liquid water that the effective one-component analysis does not capture.       
\end{abstract}
\maketitle
\section{Introduction}

Fluid suspensions of natural and technological interest are inevitably polydisperse. Their constituent particles--each of which comprise a large number of smaller atoms, molecular, or ions--exhibit distinct geometric and chemical features (size, shape, charge, etc.) and hence represent different species of a many-component mixture. As might be expected, the composition of such a mixture can strongly affect its structural, thermodynamic, and dynamic behavior.~\cite{Blum1979Polydisperse-systems, Salacuse1982Polydisperse-systems, Kotlarchyk1983Analysis-of,Briano1984Statistical-thermodynamics,Dickinson1986Polydisperse-suspensions, Barrat1986On-th-stab, Kofke1989Infinitely-polydisperse,Bolhuis1996Monte-Carlo,Phalakornkul1996Structure-and-short, vanMegen1998Measurement-of-the, Phan1998Effects-of-poly, Bartlett1999Reentrant-melting, Sear1998Phase-separation, Lacks1999Disappearances-of, Sollich2002Predicting-phase,Frydel2005-poly1,Frydel2005-poly2,Sheu2005-poly,Royall2007-poly, Sollich2007Moment-free-energies,Abraham2008Suppression-of-the, Pusey2009Hard-spheres, Sollich2011Polydispersity-induced,Berthier2011-poly,Zachary2011-poly} For example, suspensions with narrowly distributed (e.g., approximately monodisperse) particle characteristics often readily crystallize and thus can be interesting candidates for applications like colloidal self assembly.~\cite{Whitesides2002Self-assembly,Manoharan2004Building-materials} In contrast, fluids with more diverse particle populations that suppress crystallization~\cite{Barrat1986On-th-stab,Phan1998Effects-of-poly,Bartlett1999Reentrant-melting,Lacks1999Disappearances-of,Auer2001Suppression-of, Schope2007Effect-of,Pusey2009Hard-spheres,Berthier2011-poly,Zachary2011-poly} are more useful in applications which call for good glass-forming materials.

Characterizing the properties of systems with a large number of components can be a formidable technical challenge. Insufficient sampling typically limits the species-specific information that can be directly obtained via experiments and computer simulations. In fact, for suspensions with apparently ``weak" polydispersity, it is common practice to ignore particle differences altogether, adopting an effective one-component description. The latter is tantamount to assuming a fictitious pure-fluid model, whose (typically softer) interparticle interactions are those that, under equilibrium conditions, would reproduce some globally averaged behavior of the original polydisperse material.~\cite{Salgi1993Polydispersity-in,Frydel2005-poly2,Royall2007-poly,pangburn:174904,pangburn:054712} An obvious drawback of this simplification is that one generally does not know in advance whether a system's polydispersity is weak enough (as it pertains to the properties of interest) to justify use of an average description that washes out species-specific contributions. Moreover, it is difficult to assess the impact of averaging \textit{a posteriori}, especially when statistically meaningful information about the component contributions to the polydisperse material's properties is unavailable. This type of dilemma, though far from resolved, has long been appreciated and studied in colloid science. For example, it arises in considering the trade-offs between characterizing polydisperse colloidal suspensions by partial (species-specific) versus total (average) structure factors inferred from scattering data.~\cite{Blum1979Polydisperse-systems,Kotlarchyk1983Analysis-of, DAguanno1991Structural-effects,Salgi1993Polydispersity-in,Phalakornkul1996Structure-and-short, vanMegen1998Measurement-of-the,Lado1998Static-struct, Anderson2006Scattering-for, Frenkel1986Structure-factors, vanBeurten1981Polydispersity-effects, Ginoza1999Measurable-structure, Briano1984Statistical-thermodynamics}  Although a multicomponent description is ideally desired, it cannot generally be obtained from experiments without invoking a number of approximations that compromise its reliability.

In this study, we use computer simulations and exact theoretical results to examine how the static structure of a model polydisperse fluid differs when quantified based on multicomponent versus effective one-component descriptions of its pair correlations. Similar in spirit to, e.g., related work of Pangburn and Bevan,\cite{pangburn:174904,pangburn:054712} we focus exclusively on the real-space counterpart to the reciprocal-space scattering problem discussed above, the former being  especially critical for the interpretation of data from computer simulations and optical microscopy experiments\cite{Crocker1996Methods,Royall2003A-new,Varadan2003Direct,Prasad2007Confocal,iacovella2010Pair,Bevan2011149} of complex fluids. 

The model  we investigate here is a fluid of polydisperse hard-sphere particles with short-range, square-well attractions. Earlier simulations\cite{Krekelberg2007How-short-range} show that it reproduces key static and dynamic behavior of suspended colloids with polymer-mediated depletion interactions. Most notably, in agreement with experiments\cite{Eckert2002Re-entrant-Glas,Pham2002Multiple-Glassy}  it displays strikingly  anomalous dynamic behaviors at high particle concentrations near the so-called ``repulsive" glass transition (e.g., diffusivity increases and viscosity decreases with \textit{increasing} particle attractions). Interestingly, an effective one-component analysis\cite{Krekelberg2007How-short-range} suggests these behaviors emerge under conditions where the fluid also displays structural anomalies (e.g., static correlations apparently weaken with \textit{increasing} particle attractions). Of course, it is important to note that polydispersity is an important feature of this and similar models because it suppresses crystallization that would otherwise restrict access to the deeply supercooled fluid states where the anomalous behavior is most pronounced. 

Here, we use exact results to probe some of the consequences of adopting the effective one-component description for the structure of the aforementioned model in the dilute limit.   To carry out a similar analysis at high particle concentrations, we introduce Monte Carlo sampling techniques that allow us to accurately determine the $m(m+1)/2$ partial radial distribution functions as well as the free energy for a $m$ pseudo-component  [in this case, $m=60$; $m(m+1)/2=1830$] description of the model.

The main findings of this study are as follows. (1)~The multicomponent (i.e., partial) and effective one-component radial distribution functions show qualitatively different behavior as a function of the magnitude of the interparticle attractions, even for this moderately polydisperse model. This serves as a cautionary note regarding conclusions that one might otherwise draw about structure in polydisperse systems based on trends exhibited by the latter globally-averaged, static correlations. (2)~The pair-correlation contribution to the excess entropy of the polydisperse fluid based on the multicomponent description, in contrast to the estimate based on the one-component model, very closely approximates the full thermodynamic excess entropy of the system. This suggests that techniques which allow one to accurately determine the partial radial distribution functions of dense, polydisperse colloids will also furnish information key for  understanding their thermodynamic behavior. (3)~Finally, the excess entropy scaling for the shear viscosity of this polydisperse fluid is similar in some key ways to that of liquid water, another system known to exhibit anomalous thermodynamic and dynamic properties. As we show, such similarities  are not apparent if the effective one-component description is used in the excess entropy estimation for the polydisperse colloidal fluid.
\section{Methods}
\label{sec:methods}
\subsection{Model for the polydisperse fluid}\label{sec:model} We consider a previously introduced model\cite{Krekelberg2007How-short-range} comprising hard-sphere particles with short-range, square-well (SW) attractions.  The pair potential between particles $i$ and $j$ (with diameters $\sigma_i$ and $\sigma_j$, respectively) is  given by

\begin{equation}
  \label{eq:SWSRAPotential}
  \mathcal{V}_{ij}(r_{}) = \left\{
        \begin{array}{ll}
      \infty & r_{} \leq \sigma_{ij} \\
      -\epsilon & \sigma_{ij} < r \leq 1.03 \sigma_{ij}  \\
      0 & r> 1.03 \sigma_{ij}
    \end{array}
  \right.
\end{equation}Here, $r$ is the separation between particle centers, $\sigma_{ij} = (\sigma_i +\sigma_j)/2$ is the hard-core contact separation, and $\epsilon$ is  the depth of the attractive well. The particle diameters~$\sigma_{i}$ of this polydisperse fluid follow a truncated normal distribution with a mean value of $\sigma$ and a standard deviation of $0.1\sigma$.  The truncation ensures that all particle diameters  of the fluid are within 3 standard deviations of the mean ($0.7\sigma-1.3\sigma$).   This degree of polydispersity in the particle population, while still modest, is significant enough to discourage crystallization in simulations, allowing access to deeply supercooled fluid states.\cite{Krekelberg2007How-short-range} 

As discussed extensively elsewhere,\cite{Krekelberg2007How-short-range,Sciortino2002One-liquid-two-,Zaccarelli2007Colloidal} short-range attractions like those exhibited by this model can give rise to unusual dependencies of shear viscosity (which decreases) and self diffusivity (which increases) with $\epsilon/ k_{\text B} T$, where $k_{\text{B}}$ is the Boltzmann constant and $T$ is temperature. These trends capture what is observed experimentally\cite{Eckert2002Re-entrant-Glas,Pham2002Multiple-Glassy} for suspensions of hard-sphere-like colloids with polymer-mediated depletion attractions. Interestingly, an effective one-component analysis of this model's pair correlations\cite{Krekelberg2007How-short-range} suggests that the aforementioned dynamic anomalies are accompanied by a structural anomaly (i.e., increasing $\epsilon/ k_{\text B} T$ \textit{weakens} the effective pair correlations). In this work, we test whether this feature also  emerges from a multicomponent analysis.   

\subsection{Monte Carlo simulations}
 For the Monte Carlo (MC) simulations, we approximate the model introduced in Section~\ref{sec:model} by a closely related mixture of 60 pseudo-components. Particle diameters in the pseudo-component mixture exhibit continuous
polydispersity, and their pairwise interactions $\mathcal{V}_{ij}(r)$ are described by eq.~\ref{eq:SWSRAPotential}. The pseudo-components each comprise nearly identical particles (e.g., diameter differences of less than 0.01$\sigma$).  Specifically, particles with  diameters greater than $[0.7+0.01(k-1)]\sigma$ and  less than $[0.7+0.01k]\sigma$ are considered members of the $k^{th}$ pseudo-component with chemical potential $\mu_k$, where $k \in [1,60]$. In what follows, subscripts on variables refer to pseudo-component numbers rather than particle identities. We have found that  specifying the nominal mole fraction of the $k^{th}$ pseudo-component, $x_k^{\rm n}$, by the relation

\begin{equation}
        \label{eq:xjndef}
        x_k^{\rm n}=\frac{\exp \left[-\left(k-30.5\right)^2\right/200]}{\sum_{l=1}^{60}\exp \left[-\left( l-30.5\right)^2/200\right]}
\end{equation} 
results in a mixture with properties virtually identical to the model described in Section~\ref{sec:model}. As we discuss below, the pseudo-component mixture is very convenient to simulate within the semi-grand (SG) ensemble.\cite{Kofke1988Monte-Carlo}

In  our SG MC simulations of  this mixture, temperature $T$, total particle number~$N=1000$, volume~$V$ (or density~$\rho=N/$V), and all pseudo-component chemical potential differences  $\Delta \mu _{rk } =\mu _{k} -\mu _{r}$ (relative to that of an arbitrarily chosen $r^{th}$ pseudo-component reference) are held constant. In what follows, we express chemical potential differences in terms of activity ratios $\xi _{rk} = \xi _{k} / \xi _{r}$, where the activity of a particle of  the $k^{th}$ pseudo-component with molecular partition function $q_{k}$ is given by $\xi _{k} =q_{k} e^{\mu _{k} / k_{\rm B} T}$. 

In order to study a fluid with a specified pseudo-component composition $x_{k}^{{\rm n}}$ within the SG ensemble, we must first solve for the activity distribution $\xi _{rk}^{{\rm n}} $ that produces the desired $x_{k}^{{\rm n}} $ at the temperature and density of interest.  In general, this is a highly nontrivial task because $\xi _{rk}^{{\rm n}} $ is an unknown functional of $x_{k}^{{\rm n}} $.  Fortunately, this inverse problem has received considerable attention,\cite{Rutledge2001Modeling-experimental, Wilding2002Grand-canonical, Wilding2003A-nonequilibrium} and robust methods are now available for computing $\xi _{rk}^{{\rm n}} $ given $x_{k}^{{\rm n}} $.

Here, we use an efficient nonequilibrium potential refinement scheme introduced by Wilding\cite{Wilding2003A-nonequilibrium} to obtain $\xi _{rk}^{{\rm n}} $.  We take advantage of the efficient sampling of composition space facilitated by SG MC simulation techniques and allow particles to adopt \textit{any} diameter within the range $0.7 \sigma - 1.3\sigma$. 
We begin the potential refinement scheme by setting $\ln \xi _{rk}=0$ and subsequently adjust the activity distribution during an eight-stage SG MC simulation.  At regular intervals during the $i^{th}$ stage, $\ln \xi _{rk}$ is incremented by the relative difference between the instantaneous $x_{k}$ and target $x_{k}^{{\rm n}}$ discretized particle-size distributions scaled by a modification factor $\gamma _{i}$,
\begin{equation}
  \label{eq:gammaadjustment}
  \ln \xi _{rk}^{{\rm new}} =\ln \xi _{rk}^{{\rm old}} -\gamma _{i} \left(\frac{x_{k} -x_{k}^{{\rm n}} }{x_{k}^{{\rm n}} } \right).
\end{equation} A given stage terminates when the maximum relative difference $\zeta=\max \left[\left|\left(x_{k}^{\rm agg}-x_{k}^{\rm n}\right)/x_{k}^{\rm n}\right|\right]$ between the target ($x_{k}^{{\rm n}}$) and stage-averaged ($x_{k}^{{\rm agg}}$) particle size distributions drops below a tolerance of $\zeta^{*} = 0.01$. We set $\gamma _{1} =0.001$ and reduce the modification factor by a factor of two after the completion of each stage.  The activity distribution that emerges after the eighth stage is taken to be $\xi _{rk}^{{\rm n}}$.

\subsection{Static pair correlations}
Perhaps the most basic nontrivial measure of real-space structure in a homogeneous, multi-component mixture is the partial radial distribution function (PRDF), $g_{ij}(r)$. The PRDF characterizes the  frequency of various pair separations $r$ that occur between particles of ``type'' $i$ and $j$  in the fluid. Specifically,
the mean number of  particle centers of type $i$ located in a spherical shell of differential thickness $dr$ a distance $r$ away from a type $j$ particle center is $4 \pi r^2 \rho_i g_{ij}(r) dr$, where $\rho_i$ is the overall number density of particle type $i$. Note that symmetry requires $g_{ij} (r)= g_{ji}(r)$. Below, we use the labels $i$ and $j$ to indicate specific choices of the $m=60$ pseudo-components, i.e., $i,j \in [1,60]$. Note that there are $m(m+1)/2=1830$ distinct pseudo-component PRDFs in the mixture we study here, highlighting the statistical challenge associated with characterizing the structure, even at the pseudo-component level of description.   
 
The effective one-component treatment for pair correlations, on the other hand, ignores all differences between the various particle types.  Its radial distribution function $\geff(r)$ is defined such that the mean total number of particle centers located in a spherical shell of differential thickness $dr$ a distance $r$ away from other centers is $4 \pi r^2 \rho \geff (r) dr$, where $\rho$ is the total  particle number density. Expressed differently, $\geff(r)=\sum_{i=1}^M \sum_{j=1}^M x_i x_j g_{ij}(r)$. In other words, as should be expected, the effective one-component description provides a highly averaged representation of the fluid's pair correlations. This makes it easier to characterize statistically, but more challenging to interpret.
\subsection{Structural order metrics}

We examine the behavior of two different structural order metrics that characterize the degree of local translational order of a fluid based on the strength of  its pair correlations. To understand the implications of adopting the effective one-component description for our polydisperse model, we compare the value that these scalar parameters take on when the structure is described by $\geff(r)$ versus the full set of pseudo-component PRDFs, i.e., $g_{ij}(r)$, where $i,j \in [1,60]$.

The first structural metric we consider is $-\stwo/k_{\textnormal B}$,\cite{Truskett2000Towards-a-quant} where $\stwo$ comes from the leading term in the multiparticle expansion of the fluid's molar excess entropy (over ideal gas), $\sex = \stwo + s^{(3)}+\cdots$.\cite{Hernando1990Thermodynamic-pot,Samanta2001Universal-Scali}
The explicit connection between $-\stwo/k_{\textnormal B}$ and the PRDFs of a fluid mixture can be seen when the former is expressed as
\begin{equation}
  \label{s2ij}
  -\frac{\stwo}{k_\text{B}} =\sum_i \sum_j  \frac{x_{i}x_j \rho}{2} \int
  [g_{ij}(r) \ln   g_{ij}(r)-g_{ij}(r)+1]
  d{\bf{r}}
\end{equation}Note that $-\stwo/k_{\textnormal B}$ is a non-negative quantity that vanishes for an ideal gas and is considerably larger for dense liquids and glasses that exhibit stronger interparticle correlations.\cite{Truskett2000Towards-a-quant} Since $3$- and higher-body correlations are challenging to characterize in a fluid, the thermodynamic quantity $\sex$ is commonly approximated by $\stwo$ for pure fluids or binary mixtures. Here, we compute $\sex$  and $\stwo$ (at the pseudo-component level) directly from simulation. The corresponding structural order metric in the effective one-component description is given by

\begin{equation}
  \label{s2effeq}
  -\frac{\stwoeff}{k_{\textnormal B}} = \frac{\rho}{2} \int [\geff(r) \ln \geff(r) - \geff(r)+1] d\mathbf{r}
\end{equation}As discussed earlier, it was this quantity that was used in previous computer simulation studies to characterize the average strength of the interparticle correlations for the system considered here\cite{Krekelberg2007How-short-range} and related models of polydisperse colloids\cite{Mittal2006Quantitative-Li}. 

The second  structural order metric  $\tau$ studied here is defined as  \begin{equation}
  \label{eq:tau}
  \tau =y_c^{-1} \sum_i \sum_j x_i x_j  \int_0^{y_c} | g_{ij}(y) - 1 | dy
\end{equation}where $y=r \rho^{1/3}$ and $y_c$ is a cut-off value (here, we chose $y_c=4$). This measure is a straightforward generalization of a parameter introduced earlier to study the local translational order of fluid, glassy, and crystalline states of single-component materials based on their static pair correlations.\cite{Truskett2000Towards-a-quant,Errington2003Quantification-} The version of this metric for the effective one-component model is given by
 \begin{equation}
  \label{eq:tau_eff}
  \tau_{\text{eff}}=y_c^{-1}   \int_0^{y_c} | g_{\text{eff}}(y) - 1 | dy
\end{equation}where, as before, $y=r \rho^{1/3}$ and $y_c=4$.

\subsection{Thermodynamic excess entropy}   

A two-step process is used to compute the excess entropy per particle $\sex \left(\rho \sigma^3 \ ,\epsilon/k_{\rm B}T \right)$ of our pseudo-component SW fluid. In the first step, we obtain thermodynamic properties as a function of $\rho \sigma^3$ for polydisperse hard-spheres ($\epsilon/k_{\rm B}T=0$) in contact with a reservoir of particles with an ideal-gas activity distribution $\xi _{rk}^{{\rm ig}} =x_{k}^{{\rm n}} $.  In the second step, we move along a path that takes the system from $\xi _{rk}^{{\rm ig}} $ and $\epsilon/k_{\rm B}T = 0$ to $\xi _{rk}^{{\rm n}} $ and nonzero $\epsilon/k_{\rm B}T$ at constant density.

To obtain the density dependence of the entropy, we perform a multicomponent grand canonical (GC) simulation with $V = 1000 \sigma^{3}$ and $\xi _{k} =\xi _{r} \xi _{rk}^{{\rm ig}}$, with $\xi _{r} =q_{r}$. Following the strategy outlined above, we allow the individual particle diameters to take on values that span the range 0.7$\sigma$ to 1.3$\sigma$ and treat $\xi _{k} $ as a stepwise function. Transition matrix MC methods \cite{Errington2003Direct-calculat, Errington2005Direct-evaluation} are used to determine the probability $\Pi_{{\rm GC}}(N)$ of observing the system with a total number of particles \textit{N}. The density probability distribution is related to the SG partition function $\Upsilon$ and GC partition function $\Xi$ as follows\cite{Errington2005Direct-evaluation}
\begin{equation}
  \label{eq:densprobfunct}
        \Pi _{{\rm GC}} \left(N\right)=\frac{\Upsilon \left(N,\xi _{rk} ,V, T\right)}{\Xi \left(\xi _{j} ,V,T\right)} \left(\frac{\xi _{r} }{q_{r} } \right)^{N}.
\end{equation}
The relevant bridge equation $Y=-k_{{\rm B}} T\ln \Upsilon $ provides the SG potential
\begin{equation}
  \label{eq:sgpotential}
        Y\left(N\right)=-k_{{\rm B}} T\left[\ln \left(\frac{\Pi _{{\rm GC}} \left(N\right)}{\Pi _{{\rm GC}} \left(0\right)} \right)-N\ln \left(\frac{\xi _{r} }{q_{r} } \right)\right],
\end{equation}
where we have used the zero-particle limit to express the GC partition function in terms of the particle number probability distribution, $\ln \Xi =-\ln \Pi \left(0\right)$. The SG potential and Helmholtz free energy \textit{F} are linked by the relationship\cite{Kofke1988Monte-Carlo}
\begin{equation}
        \label{eq:helmholtz} 
        F\left(N\right)=Y\left(N\right)+N\sum _{k=1}^{m}\bar{x}_{k} \left(N\right)\left(\mu _{k} -\mu _{r} \right) ,       
\end{equation} 
where $\bar{x}_{k} \left(N\right)$ is the ensemble-averaged discretized particle size distribution within a system described by \textit{T}, \textit{V}, \textit{N}, and $\xi _{rk}$. For the $\epsilon/k_{\rm B}T= 0$ case, the intensive entropy \textit{s} and Helmholtz free energy \textit{f} are now
\begin{equation}
        \label{eq:intensive}
        \frac{s\left(\rho\sigma^3\ ,0\right)}{k_{{\rm B}} } =-\frac{f\left(\rho\sigma^3 ,0\right)}{k_{{\rm B}} T} =-\frac{1}{N} \ln \left(\frac{\Pi _{{\rm GC}} \left(N\right)}{\Pi _{{\rm GC}} \left(0\right)} \right)+\sum _{k=1}^{m}\bar{x}_{k} \left(N\right)\ln \left(\frac{\xi _{k} }{q_{k} } \right) ,
\end{equation}
Within the GC simulations employed here , we have $\xi _{rk}^{} =x_{k}^{{\rm n}} $, and therefore $\bar{x}_{k} \left(N\right)$ deviates from the normal particle size distribution.  For the polydisperse hard-sphere fluid ($\epsilon/k_{\rm B}T=0$), increasing $N$ causes the $\bar{x}_{k} \left(N\right)$ distribution to shift such that the populations of pseudo-components with smaller particle diameters increase. This feature of the system allows us to sample high particle number densities with relative ease.  To summarize, within the first stage of the two-step scheme employed here, we use GC simulation to obtain the density dependence of thermodynamic properties for a hard-sphere polydisperse fluid ($\epsilon/k_{\rm B}T$ = 0) subject to $\xi _{rk}^{} =x_{k}^{{\rm n}} $, and we perform just one of these simulations.

In the second step of our approach an expanded ensemble (EE) MC procedure\cite{Lyubartsev1992New-approach-to} is used to determine the difference in the thermodynamic properties of a polydisperse SW fluid with $\xi _{rk}^{{\rm n}} $ and the reference polydisperse hard-sphere fluid with $\xi _{rk}^{{\rm id}} $ at a $\rho \sigma^3$ and $\epsilon/k_{\rm B}T$ of interest. The two systems are connected through a series of subensembles described by $\lambda$, which spans from $\lambda =$ 0 to 1 in increments of $\Delta \lambda$ = 0.001 (i.e., $\lambda_i=0.001i$, where $i \in[0,1000]$). To obtain the relative entropy of the system along this path, we perform SG simulations with fixed $N = 1000$ and $\rho\sigma^3 = N\sigma^3/V$. Within subensemble \textit{i}, the activity distribution $\xi _{rk,i} $ and well depth $\epsilon_{i}/k_{\rm B}T$ are
\begin{equation}
        \label{eq:ActivityDistro}
        \ln \xi _{rk,i} =\ln \xi _{rk,i}^{{\rm id}} +\lambda _{i} \left(\ln \xi _{rk}^{{\rm n}} -\ln \xi _{rk}^{{\rm id}} \right)
\end{equation}
and
\begin{equation}
        \label{eq:welldepthk}
        \epsilon _{i} /k_{\rm B} T=\lambda _{i} \epsilon/k_{\rm B} T .
\end{equation}
Transition matrix MC methods\cite{Cichowski2005Determination-of, Errington2007Calculation-of-} are used to evaluate the probability $\Pi_{EE}\left(\lambda\right)$ of finding the system in a subensemble defined by $\lambda$. The difference in the SG potential over the entire path evaluates to
\begin{equation}
        \label{eq:deltaSGPot}
        \Delta Y=Y\left(\xi _{rk}^{{\rm n}} ,\epsilon/k_{\rm B} T \right)-Y\left(\xi _{rk}^{{\rm id}} ,0\right)=-k_{{\rm B}} T\ln \left(\frac{\Pi _{{\rm EE}} \left(\lambda =1\right)}{\Pi _{{\rm EE}} \left(\lambda =0\right)} \right).
\end{equation}
The corresponding change in the Helmholtz free energy per particle is
\begin{equation}
        \label{eq:deltahelmper}
        \Delta f\left(\rho \sigma^3 ,\epsilon/k_{\rm B} T \right)=-\frac{k_{{\rm B}} T}{N} \left[\ln \left(\frac{\Pi _{{\rm EE}} \left(1\right)}{\Pi _{{\rm EE}} \left(0\right)} \right)-N\sum _{k=1}^{m}\bar{x}_{k} \left(1\right)\ln \left(\frac{\xi _{rk}^{{\rm n}} }{q_{k} } \right)-\bar{x}_{k} \left(0\right)\ln \left(\frac{\xi _{rk}^{{\rm ig}} }{q_{k} } \right) \right]
\end{equation}
where $\bar{x}_{k} \left(0\right)$ and $\bar{x}_{k} \left(1\right)$ are the ensemble-averaged discretized particle size distributions within the $\lambda =$ 0 and 1 subensembles, respectively. By further considering the ensemble-averaged intensive configurational energies $\bar{e}\left(\lambda \right)$, we arrive at the entropy difference $\Delta s$
\begin{equation}
        \label{eq:deltaentropy}
        \frac{\Delta s\left(\rho \sigma^3 ,\epsilon/k_{\rm B} T \right)}{k_{{\rm B}} } =\frac{1}{k_{{\rm B}} T} \left[\bar{e}\left(1\right)-\bar{e}\left(0\right)-\Delta f\left(\rho \sigma^3 ,\epsilon /k_{\rm B} T\right)\right],
\end{equation}
where, in our case $\bar{e}\left(0\right)=0$.  An EE simulation is completed for each $\rho\sigma^3$ and $\epsilon/k_{\rm B} T$ of interest.  Collectively, the GC and EE simulations provide the absolute entropy
\begin{equation}
        \label{eq:absoluteentropy}
        s\left(\rho\sigma^3 ,\epsilon /k_{\rm B} T\right)=s\left(\rho\sigma^3 ,0\right)+\Delta s\left(\rho \sigma^3,\varepsilon/k_{\rm B} T \right).
\end{equation}

We now turn our attention to the ideal gas contribution to the excess entropy.  The entropy of a multicomponent ideal gas is 
\begin{equation}
        \label{eq:entropymultIG}
        \frac{s^{{\rm ig}} }{k_{{\rm B}} } =1-\ln \rho-\sum _{k=1}^{m}x_{k} \ln \left(\frac{x_{k} }{q_{k} } \right).
\end{equation}
To be consistent with the analysis above, we evaluate the expression above with $x_{k} =\bar{x}_{k} \left(1\right)$, which, of course, closely approximates $x_{k}^{{\rm n}} $.  Finally, we arrive at the excess entropy
\begin{equation}
        \label{eq:excessentropy}
        s^{{\rm ex}} =s-s^{{\rm ig}} .
\end{equation}
We note that the algorithm outlined here is rather general and we expect it to prove useful for computing the free energy of a wide range of polydisperse fluids.

 \begin{figure}[t]
  \centering
  \includegraphics[width=2.75in]{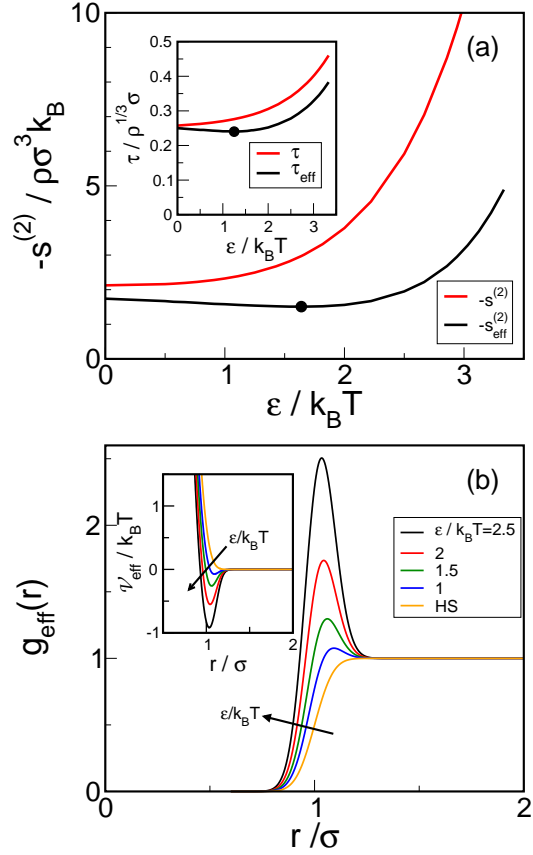}
  \caption{Effect that the magnitude of the interparticle attraction $\epsilon/k_{\rm B} T$ has on the static structure of the polydisperse SW fluid  described in the text in the dilute ($\rho \sigma^3 \rightarrow 0$) limit. (a) [Main panel] Comparison of the reduced two-body excess entropy, $-\stwo/\rho \sigma^3 k_{\rm B}$ (identical to $-\sex/\rho \sigma^3 k_{\rm B}$ in this limit), computed using eq.~\ref{eq:apps2ij}, as well its effective one-component counterpart, $-\stwoeff \rho \sigma^3 k_{\rm B}$, calculated using eq.~\ref{s2effeq} and \ref{eq:appgeff}.  [Inset]  Comparison of the reduced structural order metric, $\tau/\rho^{1/3}\sigma$, computed using eq.\pageref{eq:apptau}, as well its effective one-component counterpart, $\taueff /\rho^{1/3}\sigma$, calculated using eq.~\ref{eq:tau_eff} and \ref{eq:appgeff}. Effective one-component quantities spuriously indicate the presence of structural anomalies (attractions apparently weaken structure) for energies $\epsilon/k_{\rm B} T$ less than the values indicated by the black dots. (b) [Main panel] The effective one-component radial distribution function $\geff(r)$, computed using eq.~\ref{eq:appgeff}, plotted versus reduced interparticle separation $r/\sigma$. [Inset] The interparticle potential, $\mathcal{V}_{\rm eff}(r)=-k_{\rm B} T \ln \geff (r)$, of the effective one-component fluid.}
  \label{fig:zerodens}
\end{figure}
\begin{figure}
        \centering
        \includegraphics[width=3.5in]{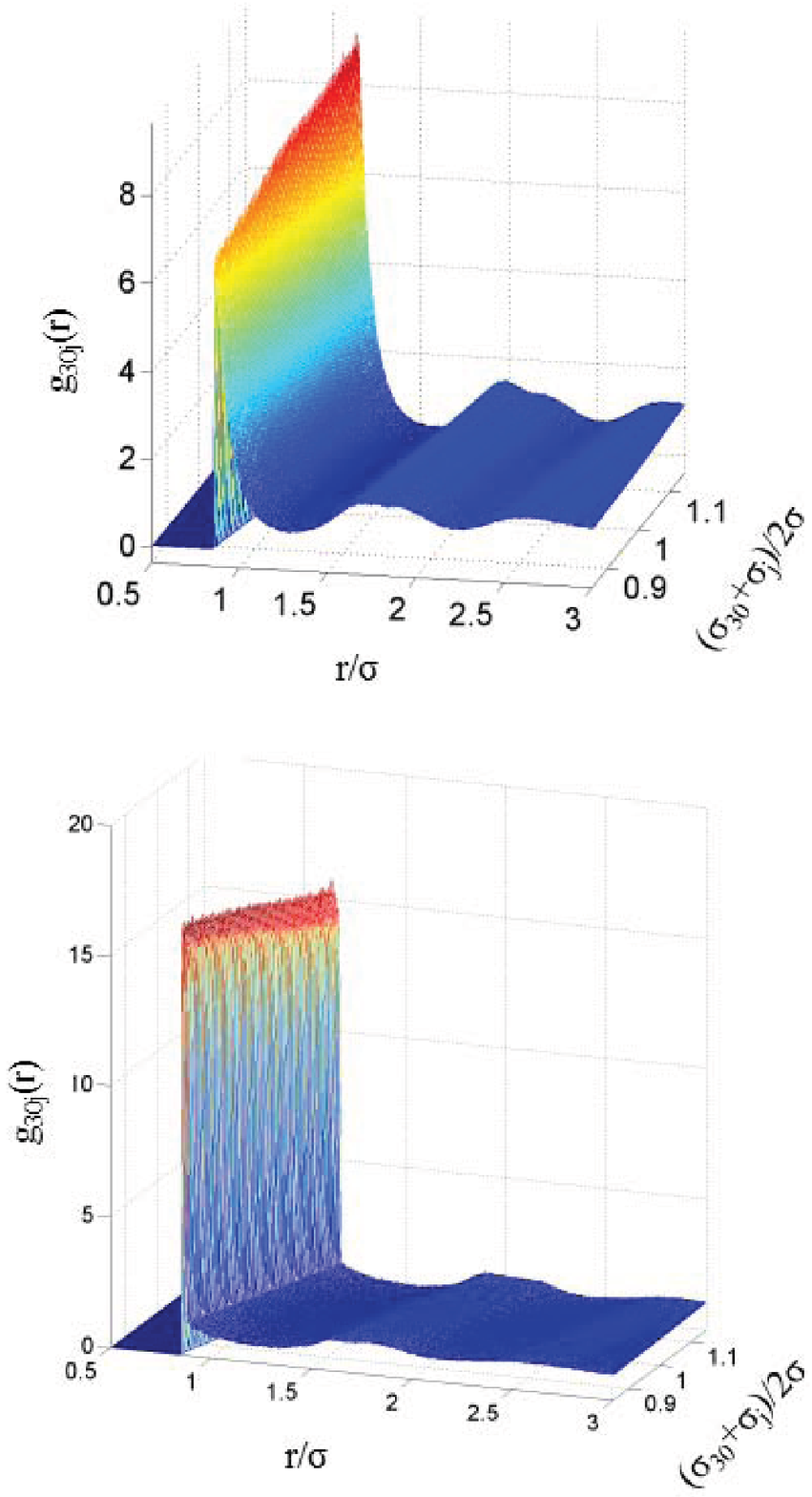}
        \caption{\small{Partial radial distribution functions $g_{30j}(r)$ describing correlations between the $30^{th}$ and the $j^{th}$ pseudo-components ($j \in[1,60]) $ of the polydisperse SW fluid  obtained via MC simulations. Results are shown as a function of reduced interparticle separation $r/\sigma$ and interaction diameter $(\sigma_{30}+\sigma_j)/2$. Color (red-to-blue) corresponds to magnitude of $g_{30j}(r)$ (high-to-low). Data are for a reduced density of $\rho \sigma^3=1.05$ and interparticle attractions of (top panel) $\epsilon/k_{\rm B}T =0$ (hard-sphere limit) and (bottom panel) $\epsilon/k_{\rm B}T =2.5$.}} 
        \label{fig:HS3D}
\end{figure}


\begin{figure}
        \centering
        \includegraphics[width=3in]{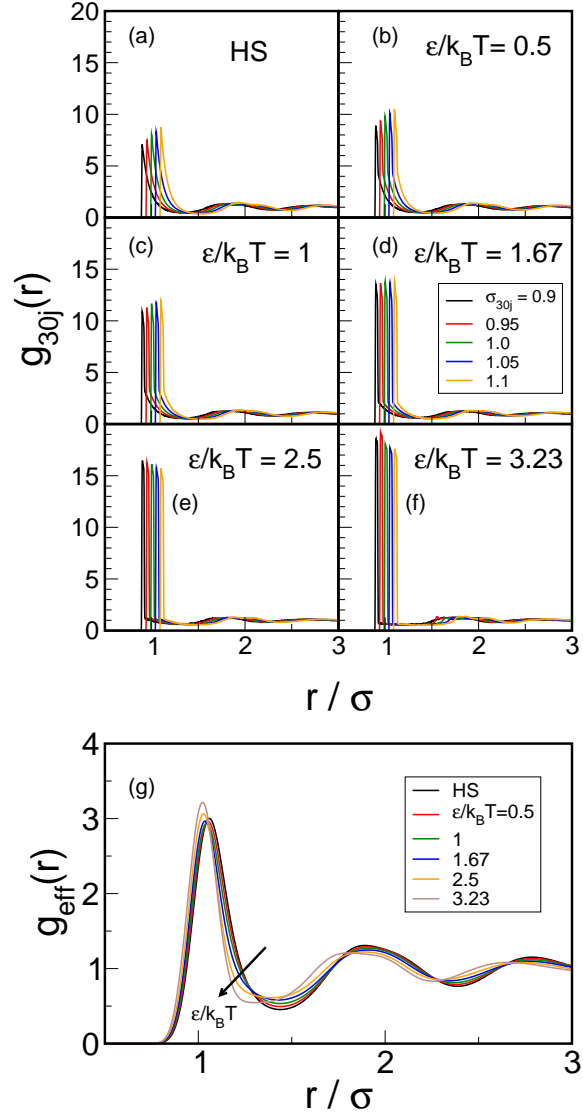}
        \caption{Effect that magnitude of the interparticle attraction $\epsilon/k_{\rm B}T$ has on the static pair correlations of the polydisperse SW fluid at a reduced density of $\rho \sigma^3=1.05$. Data obtained via MC simulations. (a)-(f) Partial radial distribution functions $g_{30j}(r)$ between $30^{th}$ and the $j^{th}$ pseudo-components for $j=10$, $20$, $30$, $40$, and $50$ (with corresponding characteristic interaction diameters $\sigma_{30j}=0.90$, $0.95$, $1.0$, $1.05$ and $1.10$, respectively) and various values of  $\epsilon/k_{\rm B}T$. (g) Radial distribution function for the effective one-component description of the fluid, $\geff(r),$ as a function of reduced particle center separation~$r/\sigma$ for conditions used to generate the data for  panels (a)-(f).}
        \label{fig:phic55gofrs}
\end{figure}


\begin{figure}[t]
  \centering
  \includegraphics[width=2.75in]{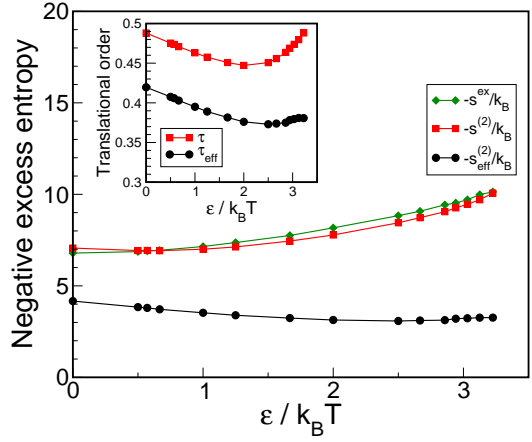}
  \caption{Effect that magnitude of interparticle attraction  $\epsilon/k_{\rm B}T$ has on different structural order metrics  for the polydisperse SW fluid at a reduced density of $\rho \sigma^3=1.05$ Data obtained via MC simulations. [Main panel] Negative  thermodynamic excess entropy, $-\sex/k_{\rm B}$ (diamonds), and its two-body approximation based on multicomponent, $-\stwo/k_{\rm B}$ (squares), and effective one-component, $-\stwoeff/k_{\rm B}$ (circles), descriptions of the system.  [Inset] Translational structure metrics based on the effective one-component, $\taueff$ (circles), and multicomponent, $\tau$ (squares), descriptions of the system. Curves are guides to the eye. }
  \label{fig:highdens}
\end{figure}

\section{Results and discussion}

The first results that we consider pertain to the behavior  of a dilute ($\rho \rightarrow 0$) polydisperse fluid of SW particles that interact through the potential given by eq.~\ref{eq:SWSRAPotential} and that exhibit a Gaussian distribution of particle diameters (with mean $\sigma$ and standard deviation $\sigma/10$). This is an interesting  case to study because--while simple enough to allow for  an exact theoretical  description of the polydisperse fluid's static properties--it is also rich enough to illustrate some of the key consequences of adopting the effective one-component description.
Analytical expressions for the structural quantities discussed below can be found in the Appendix.

Fig.~\ref{fig:zerodens} shows how the strength of the static interparticle correlations of the dilute fluid depends on the magnitude of the interparticle attraction~$\epsilon/k_{\rm B} T$. Two levels of structural characterization are considered: one that takes into account the full set of PRDFs (i.e., the exact multicomponent description), and one based on the effective one-component structure. The key point of Fig.~\ref{fig:zerodens}a is that the structural order metrics that properly account for the  polydisperse nature of the fluid ($-\stwo/\rho \sigma^3 k_{\rm B}$ and $\tau/\rho^{1/3}\sigma$) show qualitatively different behavior than their effective one-component counterparts ($-\stwoeff/\rho \sigma^3 k_{\rm B}$ and $\taueff/\rho^{1/3}\sigma$).  The former display the expected monotonic behavior; i.e., static correlations strengthen with increasing magnitude of the interparticle attractions. In fact, in the Appendix, we demonstrate that this monotonic trend must always be observed in the dilute limit for this system, independent of the width of the diameter distribution. In contrast, the effective one-component quantities both spuriously indicate the presence of a structural anomaly; i.e., conditions where static correlations apparently weaken with increasing interparticle attractions. 

Why do $-\stwoeff/\rho \sigma^3 k_{\rm B}$ and $\taueff/\rho^{1/3}\sigma$ predict qualitatively different behavior than $-\stwo/\rho \sigma^3 k_{\rm B}$ and $\tau/\rho^{1/3}\sigma$, respectively? The basic mathematical reason is that the structural metrics are nonlinear functionals of the PRDFs. Thus, their behaviors can depend sensitively on whether the latter ``inputs" are treated in a multicomponent  or a pre-averaged (effective one-component) manner. From a physical perspective,  as illustrated in Fig.~\ref{fig:zerodens}b, averaging the PRDFs to obtain $\geff(r)$  softens the apparent pair correlations in the fluid, as well as the effective interactions between the particles, $\mathcal{V}_{\rm eff}(r)=-k_{\rm B}T \ln \geff(r)$. As a result, the effective one-component  metrics significantly  underestimate the effect that the attractions have on static structure. For example, the magnitude of $-\stwoeff/\rho \sigma^3 k_{\rm B}$ is much smaller than that of $-\stwo/\rho \sigma^3 k_{\rm B}$ (which is equal to $-\sex/\rho \sigma^3 k_{\rm B}$ in this dilute limit) for $\epsilon/k_{\rm B} T>1$, illustrating that one cannot readily estimate the thermodynamic excess entropy from the effective one-component approximation.  

In order to compare this behavior to that of concentrated SW fluids, we
analyze results from our MC simulations. Fig.~\ref{fig:HS3D} displays 60 of the 1830 PRDFs that we collected for the polydisperse SW fluid described in Section~\ref{sec:methods} at a reduced density $\rho \sigma^3=1.05$ (particle packing fraction of 0.567) and two extreme levels of interparticle attractions: $\epsilon/k_{\rm B} T=0$~(top panel of Fig.~\ref{fig:HS3D}) and $\epsilon/k_{\rm B} T=2.5$ (bottom panel  of Fig.~\ref{fig:HS3D}). The functions presented in Fig.~\ref{fig:HS3D}, labeled $g_{30j}(r)$, characterize the partial pair correlations between  particles with diameters close to the median size (the $30^{th}$ pseudo-component) and particles of other sizes (the $j^{th}$ pseudo-component with $j \in[1,60])$. In the case of $\epsilon/k_{\rm B} T=0$ (i.e., the hard-sphere limit), there is a clear relationship between the contact value $g_{30j}(r=\sigma_{30j}$) and the contact diameter $\sigma_{30j}=(\sigma_{30}+\sigma_{j})/2$; larger particles have higher probabilities of contact configurations.  This is the expected trend and is a classic manifestation of the depletion effect.\cite{Asakura1954On-Interaction-,Reiss1959Statistical-mec} Contact configurations of big spheres are entropically favorable  because they increase the volume accessible to smaller spheres. A weak splitting of the second peak of the PRDFs is also apparent--as is commonly observed for dense hard-sphere fluids\cite{Truskett1998Structural-prec}--the precise shape of which depends on the identities of the participating pseudo-components. Interestingly, for $\epsilon/k_{\rm B} T=2.5$, the aforementioned entropic packing preferences have all but vanished. The strong attractions in this limit dominate the pair correlations, and there are only very minor differences between PRDFs involving different pseudo-components.  The main point is that the  sensitivity of the distribution of PRDFs to  $\epsilon/k_{\rm B} T$ suggests that the pre-averaging required for the effective one-component description may also be problematic for this high-density system, as it was for the dilute case.

The data of Fig.~\ref{fig:phic55gofrs} clarify this point. Specifically, panels~\ref{fig:phic55gofrs}a-\ref{fig:phic55gofrs}f illustrate how 5 characteristic PRDFs for this system--$g_{30j}(r)$  for $j=10$, $20$, $30$, $40$, and $50$--evolve with increasing strength of the interparticle attraction. The pronounced changes  of these functions  [e.g., the contact values  $g_{30j}(r=\sigma_{30j}$) triple in magnitude with increasing  $\epsilon/k_{\rm B} T$] should be contrasted with the behavior of the corresponding effective one-component quantity $\geff(r)$ shown in Figure~\ref{fig:phic55gofrs}g, which not only looks qualitatively different (considerably softer  than the PRDFs) but also is remarkably insensitive to the value of $\epsilon/k_{\rm B} T$ considered. In other words, the averaging involved in forming $\geff(r)$ has the effect of hiding some significant changes that attractions have on the static structure of the polydisperse fluid.  Similar-if less dramatic--effects emerge when adopting  coarse-grained structural descriptions of other types of materials (e.g., center-of-mass versus monomer-level pair-correlations in fluids of chain-like molecules\cite{Goel2008Excess-entropy}). 

The consequences of the  aforementioned averaging become clearer in Fig.~\ref{fig:highdens} where we directly compare how the structural order metrics--based on multi-component versus one-component descriptions of the systems--depend on   $\epsilon/k_{\rm B}T$. The first point to note is that, as in the dilute fluid, the  metrics $-\stwo/k_{\rm B}$ and $-\stwoeff/k_{\rm B}$ show different behavior, both quantitatively and qualitatively. The former provides an excellent approximation to the thermodynamic excess entropy $-\sex/k_{\rm B}$, while the latter is too small by a factor of between 1.8 and 3.5 (depending on $\epsilon/k_{\rm B}T$). Also, as observed in the dilute case, the effective one-component quantity reports that the fluid is structurally anomalous (attractions apparently weaken pair correlations) across  a wide range of $\epsilon/k_{\rm B}T$. In contrast, $-\stwo/k_{\rm B}$ shows normal behavior (attractions strengthen correlations) except for a very narrow range of conditions at low values of $\epsilon/k_{\rm B}T$ where a weak structural anomaly is present. The quantity $-\sex/k_{\rm B}$ displays normal behavior for all $\epsilon/k_{\rm B}T$. Similar qualitative trends are apparent when comparing  $\tau$ and $\taueff$. The latter underestimates the magnitude of the structural order, and it overestimates the range of conditions where structural anomalies occur.

\begin{figure}
        \centering
        \includegraphics[clip]{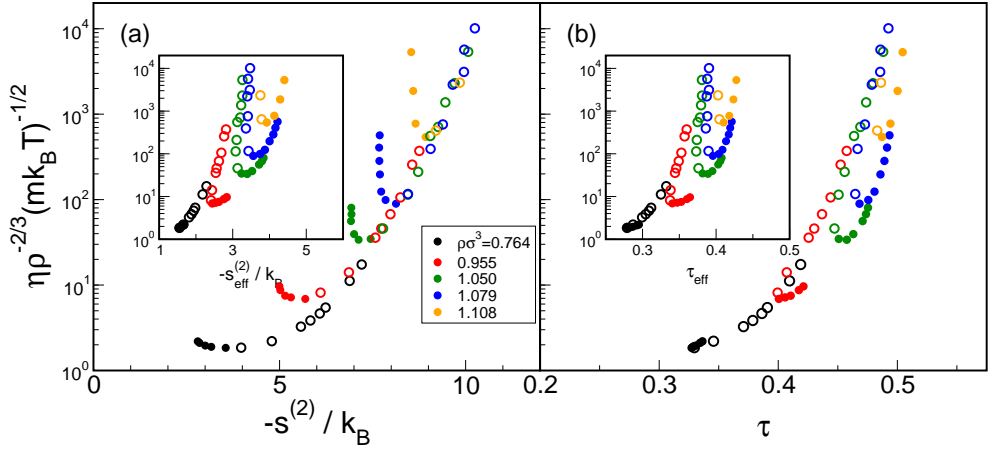}
        \caption{Reduced shear viscosity $\eta \rho^{-2/3}(m k_{\rm B}T)^{-1/2}$ plotted versus various structural order metrics based on static pair correlations for the polydisperse SW fluid: (a) [main panel] negative two-body excess entropy, $-\stwo/k_{\rm B}$ and [inset] its effective one-component counterpart, $-\stwoeff/k_{\rm B}$; (b) [main panel] translational structure metric, $\tau$,  and [inset] its effective one-component counterpart, $\taueff$. Structural data were obtained via the MC simulations of this study, and the dynamic data were extracted from the earlier molecular dynamics investigation of  Krekelberg et al.\cite{Krekelberg2007How-short-range}. Open symbols represent data for $\epsilon/k_{\rm B}T>2$ and filled symbols for   $\epsilon/k_{\rm B}T<2$. For the latter conditions, shear viscosity anomalously increases with $\epsilon/k_{\rm B}T$ (see Fig. 2 of ref.~\onlinecite{Krekelberg2007How-short-range})}
       \label{fig:scaling}
\end{figure}

As a final point, we examine the implications of the one-component
analysis for scaling relationships between static structure and dynamics. In particular, we plot in Fig.~\ref{fig:scaling} reduced shear
viscosity $\eta \rho^{-2/3}(m k_{\rm B}T)^{-1/2}$ (here $\eta$ shear viscosity and $m$ is particle mass) for 
the polydisperse SW fluid as a function of  the various structural order metrics analyzed above.  The shear viscosity data we use for this plot was extracted from the earlier molecular dynamics investigation of  Krekelberg et al.\cite{Krekelberg2007How-short-range}. As Rosenfeld\cite{Rosenfeld1977Relation-betwee,Rosenfeld1999A-quasi-univers} originally observed, the non-dimensionalized form $\eta \rho^{-2/3}(m k_{\rm B}T)^{-1/2}$  is strictly a single-valued function of $-\sex/k_{\rm B}$ for a fluid of classical particles interacting via a pair potential of the  inverse-power-law form and obeying Newton's equations of motion. For fluids with more complex structures and interactions, the scaling can only be expected to approximately hold.

The key point of Fig.~\ref{fig:scaling} is that one sees qualitatively different trends for the scaling relation depending on whether the multicomponent ($-\stwo/k_{\rm B}$ or $\tau$) or effective one-component quantities $(-\stwoeff/k_{\rm B}$ or $\taueff$) are used. When the former are adopted, one finds that a single scaling curve describes all data for which $\epsilon/k_{\rm B}T>2$ (i.e., the data for which $\eta$ follows the ``dynamically normal" trend of increasing with $\epsilon/k_{\rm B}T$; see Fig. 2 of ref.~\onlinecite{Krekelberg2007How-short-range}). Only dynamically anomalous state points with $\epsilon/k_{\rm B}T<2$ (i.e., for which $\eta$ decreases with increasing $\epsilon/k_{\rm B}T$) deviate from the scaling curve.  This is  analogous to the excess entropy scaling behavior observed in recent simulations of liquid water.\cite{Chopra2010On-the-use} Specifically, deviations from excess entropy scaling were also observed   for dynamically anomalous state points  (where water's relaxation times decrease with \textit{increasing} density). As can be seen from the insets of Fig.~\ref{fig:scaling} (also in the data of ref.~\onlinecite{Krekelberg2007How-short-range}), different trends emerge when the effective one-component quantities $-\stwoeff/k_{\rm B}$ or $\taueff$ are adopted in the analysis, which obscure the aforementioned qualitative connection between the scaling behaviors of two different types of fluids that exhibit dynamic anomalies.
\section{Conclusions}

This investigation highlights some significant problems that arise when characterizing real-space, static pair correlations of moderately polydisperse colloidal fluids by a commonly-used effective one-component approach. Specifically, our analysis of a model fluid that mimics the experimental behavior of colloids with short-range attractions shows that trends based on the one-component treatment are qualitatively inconsistent with those of the full multicomponent description; e.g., the former spuriously indicates that pair correlations weaken with increasing attractions over a wide range of conditions. A key problem associated with the effective one-component analysis is an  artificial, polydispersity-induced ``softening'' of the apparent static correlations--due to averaging over different particle sizes--which in turn causes the entropic consequences of structural ordering to be significantly underestimated. As an example of the implications of this issue, we demonstrate how excess entropy scalings for the shear viscosity of our model fluid show expected behavior only when the full multicomponent structural analysis is employed. Finally, we show that characterization of the partial radial distribution functions in  concentrated suspensions of model colloidal fluid allows for an accurate estimate of the thermodynamic excess entropy, a quantity that cannot be reliably estimated from an effective one-component analysis.

What is the maximum level of polydispersity (e.g., variance in the particle diameter distribution) for which the aforementioned effective one-component description can accurately capture the structural properties of multicomponent fluids? Furthermore, how does the answer depend on the nature of the polydispersity? At present, the answers to these  important questions  are unknown. We are currently studying both issues, and we will report our findings in a future publication.  

 On the theoretical and methods fronts, the present study demonstrates that both analytical results from statistical mechanics and Monte Carlo simulations can provide insights into some of the key polydispersity-related issues for the structural characterization of model colloidal fluids. Another important open question is how to carry out a similar multicomponent analysis for data obtained via, e.g.,  optical microscopy experiments, where one is inevitably limited by inadequate sampling.  We are currently exploring the prospects of using maximum likelihood estimation techniques together with liquid-state theory for accomplishing this, and we will report our findings in a subsequent publication.  
\section*{Acknowledgements}

We thank Prof. Stuart Rice and Prof. C. P. Royall for their helpful input.  T.M.T. acknowledges support of the
Welch Foundation (F-1696) and the National Sciece Foundation (CBET-1065357).
J. R. E. acknowledges financial support of the National Science Foundation (CBET-0828979).  M.J.P. acknowledges the support of the Thrust 2000 - Harry P.
Whitworth Endowed Graduate Fellowship in Engineering. The Texas
Advanced Computing Center (TACC), the University at Buffalo Center for Computational Research, and the Rensselaer Polytechnic Institute Computational Center for Nanotechnology Innovations provided computational resources for
this study.
\section*{Appendix: Exact results for low particle density}
\label{app}

 In the limit of low particle density ($\rho \rightarrow 0$), some implications of adopting the effective one-component description can be worked out analytically for a continuously polydisperse SW fluid with pair interaction $\mathcal{V}_{ij}(r)$,\begin{equation}
  \label{eq:appenPotential}
  \mathcal{V}_{ij}(r_{}) = \left\{
        \begin{array}{ll}
      \infty & r_{} \leq \sigma_{ij} \\
      -\epsilon & \sigma_{ij} < r \leq (1+\lambda) \sigma_{ij}  \\
      0 & r> (1+\lambda) \sigma_{ij}
    \end{array}
  \right.
\end{equation}and a Gaussian distribution of particle  diameters with mean $\sigma$ and variance $\upsilon^{2}\sigma^{2}$, i.e., $p(\sigma_{i})=(2\pi\upsilon^{2}\sigma^{2})^{-1/2}\exp[-(\sigma_{i}-\sigma)^{2}/2\upsilon^{2}\sigma^{2}]$. Note that, for this system, the  probability distribution for the interaction diameter,    $f(\sigma_{ij})$ is a Gaussian with mean $\sigma$ and variance $\nu^2\sigma^2$/2. In the low density limit, the PRDFs are known exactly, i.e.,

\begin{equation}
  \label{eq:appenPRDF}
  g_{ij}(r) = \left\{
        \begin{array}{ll}
      0 & r_{} \leq \sigma_{ij} \\
      e^{\epsilon/k_{\rm B}T} & \sigma_{ij} < r \leq (1+\lambda) \sigma_{ij}  \\
      1 & r> (1+\lambda) \sigma_{ij}
    \end{array}
  \right.
\end{equation}       
Substituting eq.~\ref{eq:appenPRDF} into the continuous polydispersity limit of eq.~\ref{s2ij}, $-\stwo/k_{\text{B}} = (\rho/2)\int d \sigma_{ij} f(\sigma_{ij})  \int d{\bf{r}}[g_{ij}(r) \ln   g_{ij}(r)-g_{ij}(r)+1]$, noting that $\langle \sigma_{ij}^3\rangle=\sigma^3(1+3\nu^2/2)$, and evaluating yields
\begin{equation}
  \label{eq:apps2ij}
  -\frac{\stwo}{ k_\text{B}\rho\sigma^3} =\frac{2\pi}{3} \left\{1+\frac{3\nu^2}{2}
\right\}\left\{1+\left[(1+\lambda)^3-1\right]\left[e^{\epsilon/k_{\rm B}T}\left(\frac{\epsilon}{k_{\rm B}T}-1\right)+1\right]\right\}\end{equation}The right-hand-side of eq.~\ref{eq:apps2ij} is equal to $B/\sigma^{3}+[d(B/\sigma^{3})/d\ln T$], where $B$ is the second virial coefficient of the fluid. Thus, to leading order in density, we also have $\stwo=\sex$, as expected. \ Likewise,  substituting eq.~\ref{eq:appenPRDF} into the continuous polydispersity limit of eq.~\ref{eq:tau}, $\tau =y_c^{-1} \int d \sigma_{ij} f(\sigma_{ij})\int_0^{y_c}  dy| g_{ij}(y) - 1 |$, and evaluating  gives the following expression \begin{equation}
  \label{eq:apptau}
  \frac{\tau}{\rho^{1/3}\sigma} =\frac{1}{y_{c}}\left\{1+\lambda\left(e^{\epsilon/k_{\rm B}T}-1\right)\right\}
\end{equation}
As can readily be seen, the right-hand sides of both eq.~\ref{eq:apps2ij} and \ref{eq:apptau} are monotonically increasing functions of $\epsilon/k_{\rm B}T$. In other words, no structural anomalies appear in the dilute SW fluid according to the exact,  multicomponent analysis. This is consistent with the recently reported low-density behavior~\cite{Krekel-struct-anom} of  a related single-component model.

The effective one-component  radial distribution $\geff(r)=  \int d \sigma_{ij} f(\sigma_{ij})g_{ij} (r)$ can also be obtained  from knowledge $f(\sigma_{ij})$ and $g_{ij} (r)$ of eq.~\ref{eq:appenPRDF}; it is given by 
\begin{equation}
\label{eq:appgeff}             
g_{{\rm eff}}\left(r\right)=\frac{1}{2}\left[1+\left(1-e^{\epsilon/k_{\rm B}T }\right){\rm erf}\left\{\frac{r/\sigma-(1+\lambda )}{v(1+\lambda )}\right\}+e^{\epsilon/k_{\rm B}T }{\rm erf}\left\{\frac{r/\sigma-1}{v}\right\}\right]
\end{equation}
Substitution of eq.~\ref{eq:appgeff} into eq.~\ref{s2effeq} and \ref{eq:tau_eff}  allows determination of $-\stwoeff$ and $\taueff$, respectively. Note that these quantities, unlike  $-\stwo$ and $\tau$, show a minimum  at intermediate values of $\epsilon/k_{\rm B}T$, giving the spurious impression that  pair correlations anomalously   weaken with attractions at low values of $\epsilon/k_{\rm B}T$. 

As a final point, we note that the single-component system with structure equivalent to that indicated by the  effective one-component analysis will  have the following pair potential, $\mathcal{V}_{\rm eff}(r)=-k_{\rm B} T \ln \geff (r)$.  Polydispersity makes this potential considerably softer than the SW interaction between any two particles in the polydisperse fluid.

        
\end{document}